\documentclass{caosp}

\usepackage{graphicx}

\articleNo{BP16}
\pubyear{2014}
\volume{43}
\volnumber{3}
\firstpage{1}
\received{November 4, 2013}
\accepted{January 27, 2014}

\begin{document}

\htitle{Medium-resolution echelle spectroscopy of pulsating variables\ldots}
\hauthor{J.\,Kov\'{a}cs, B.\,Cs\'{a}k,  B.\,Cseh, L.\,Szabados, Gy.M.\,Szab\'{o},  \'{A}.\,D\'{o}zsa,  L.L.\,Kiss and  I.\,Jankovics}

\title{Medium-resolution echelle spectroscopy of pulsating variables and exoplanet host stars with sub-meter telescopes}

\author{
        J.\,Kov\'{a}cs\inst{1}
      \and
        B.\,Cs\'{a}k\inst{1}
      \and
        B.\,Cseh\inst{1}
      \and
        L.\,Szabados\inst{2}
      \and
        Gy.M.\,Szab\'{o}\inst{1,2}
      \and
        \'{A}.\,D\'{o}zsa\inst{1}
      \and
        L.L.\,Kiss\inst{2}
      \and
        I.\,Jankovics\inst{1}
       }

\institute{
           ELTE Gothard Astrophysical Observatory, Szombathely, Hungary
         \and 
           Konkoly Observatory of Hungarian Academy of Sciences, Budapest, Hungary
          }

\date{November 4, 2013}

\maketitle

\begin{abstract}
Here we present two of our interesting results obtained over the last 18
months from spectroscopic monitoring of binary pulsating stars and exoplanet
host stars.  Our investigations are very promising by demonstrating that
modern fiber-fed spectrographs open a whole new chapter in the life of small
national and university observatories.
\end{abstract}

\section{Introduction}

Medium-resolution echelle spectroscopy that is capable of $\pm 50~\mathrm{m\,s}^{-1}$ radial velocity precision was first introduced to Hungarian observatories in early 2012. The instrument, the commercially available eShel echelle spectrograph of the French Shelyak Instruments (Thizy \& Cochard, 2011) offers an exceptional light throughput that allows better than $100~\mathrm{m\,s}^{-1}$ radial velocities for 9--11th magnitude stars with typical integrations of 20--60 minutes. With this precision a range of ground-based spectroscopic support work has become possible.

\section{Telescopes, spectrograph \& observations}

The main light gathering instrument for the spectrograph is the 0.5m RC telescope of the Gothard Astrophysical Observatory (GAO), but thanks to its transportability it has regularly been used on the 1m RCC telescope at Piszk\'estet\H{o} Mountain Station of the Konkoly Observatory (PO).
The fiber injection and guiding unit of the spectrograph is attached to the
Cassegrain focus of the telescopes, while the spectrograph is located in a thermally isolated room. The light from the object and ThAr fibers is collimated and projected to an R2 echelle grating, which is typically used between orders \#29 and \#56 separated by a prism.
The entire process of data acquisition is fully computer controlled.
The system is surprisingly compact and lightweight and it is easy to carry and install on different telescopes.
Over the last 1.5 years more than $16\,000$ scientific frames including about $4\,700$ stellar spectra of about 180 stars have been recorded.

\section{Exoplanet host stars}

Doppler observations of exoplanet systems have been a very expensive technique, mainly due to the high costs of high-resolution stable spectrographs. However, recent advances in instrumentation enable affordable Doppler planet detections with surprisingly small optical telescopes that have traditionally been neglected for this kind of studies.
Our target list includes well-known bright exoplanet host stars, as well as somewhat fainter objects from WASP and HATNet projects, e.g. HAT-P-2 (Bakos, 2007).

\begin{figure}
\vspace*{-5.mm}
\centerline{\includegraphics[width=0.6\textwidth]{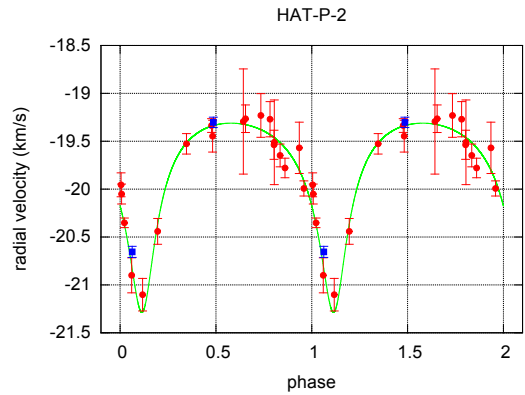}}
\vspace*{-2.mm}
\caption{A model RV curve of exoplanet host star HAT-P-2 based on photometric observations overplotted with our (circles) and Mercator/HERMES (squares) radial velocity measurements. Note that the accuracy of our data is sometimes better than $100~\mathrm{m\,s}^{-1}$, which  clearly shows the applicability of our technique for detection exoplanets with a relatively small RV amplitude.}
\label{HAT-P-2}
\vspace*{-7.mm}
\end{figure}

\vspace*{-2.mm}
\section{Pulsating variables}
Photometry of pulsating variables is a traditionally successful research field of the Hungarian astronomy, but over the last century no spectroscopic observations were carried out in our observatories. With the acquisition and installation of a medium-resolution mobile echelle spectrograph our aim was to change this situation, especially in the case of Cepheids.

Classical Cepheid variables are well-known primary distance indicators owing to the $P$-$L$ relationship. Companions to Cepheids, however, complicate its applicability for distance determination (Szabados {\it et al.}, 2013).
Binaries among Cepheids are not rare at all: their frequency exceeds 50\% for the brightest Cepheids, while among the fainter Cepheids an observational selection effect encumbers revealing binarity (Szabados, 2003).
In the case of pulsating variables, spectroscopic binarity (SB) manifests itself in a periodic variation of the $\gamma$-velocity. In practice, the orbital radial velocity variation of the Cepheid component is superimposed on the RV variations of pulsational origin.

The very first result of this work is pointing out the SB nature of the bright Galactic Cepheid V1344 Aql by analysing old and our new radial velocity measurements
(Szabados {\it et al.}, 2014). Two earlier data sets (Balona, 1981 and ArenalloFerro, 1984) already imply a slight shift between the annual mean velocities. Supplemented with our new data, the merged phase diagram of all RV observations clearly shows a vertical shift between the mean values valid for the early 1980s and 2012. Though the difference is small, quality of the data and the identical treatment is a guarantee that the shift is real.

Regular monitoring of the RVs of a large number of Cepheids will be instrumental in finding more long-period SBs among them. Data to be obtained
within the Gaia astrometric space mission will certainly result in revealing many new SBs among Cepheids brighter than 13--14th magnitude. As to V1344 Aql, we keep this Cepheid among our targets.

\vspace*{-3.0ex}
\begin{figure}
\centerline{\includegraphics[width=0.5\textwidth,clip=]{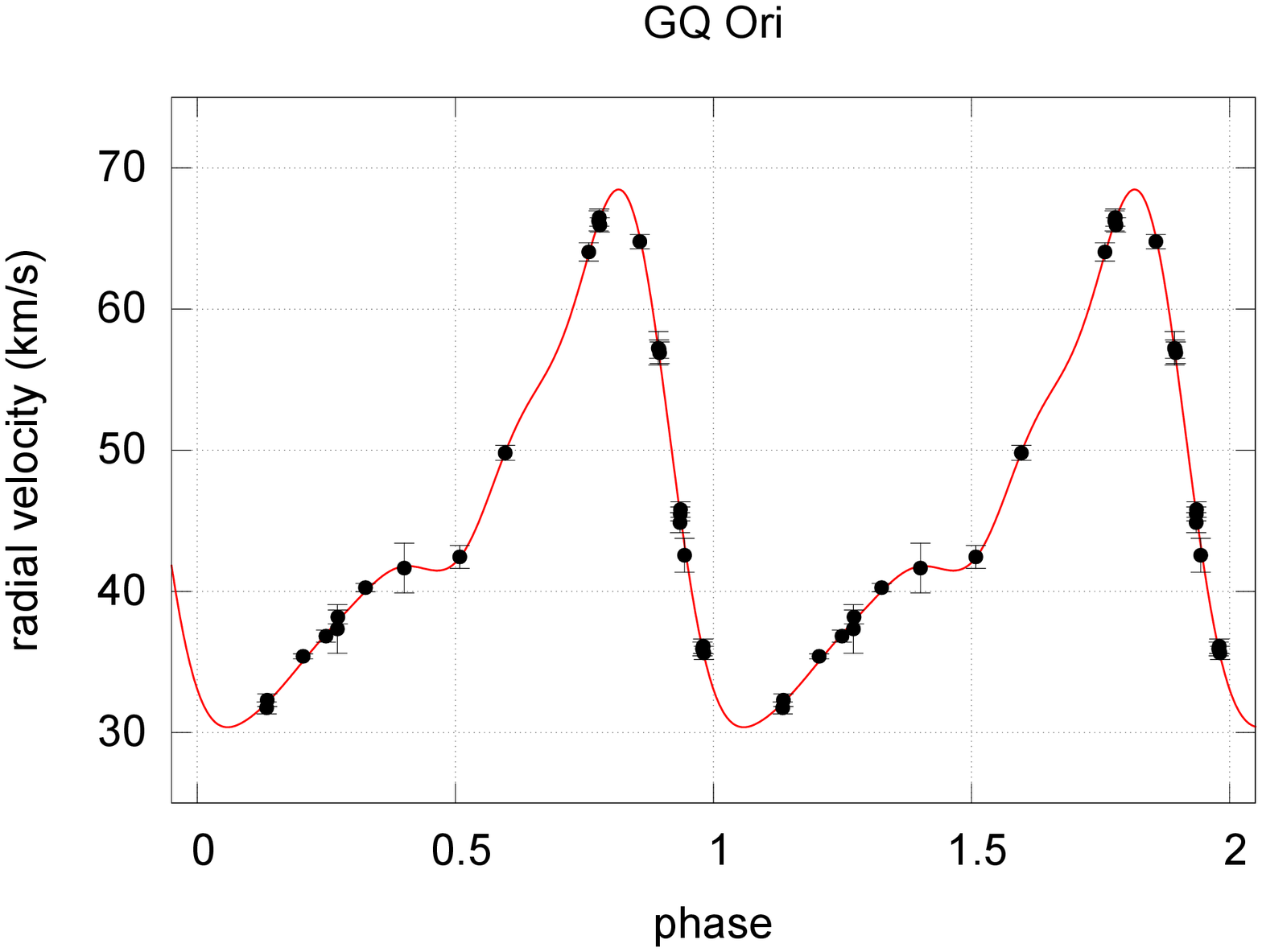}
 \hfill
 \includegraphics[width=0.5\textwidth,clip=]{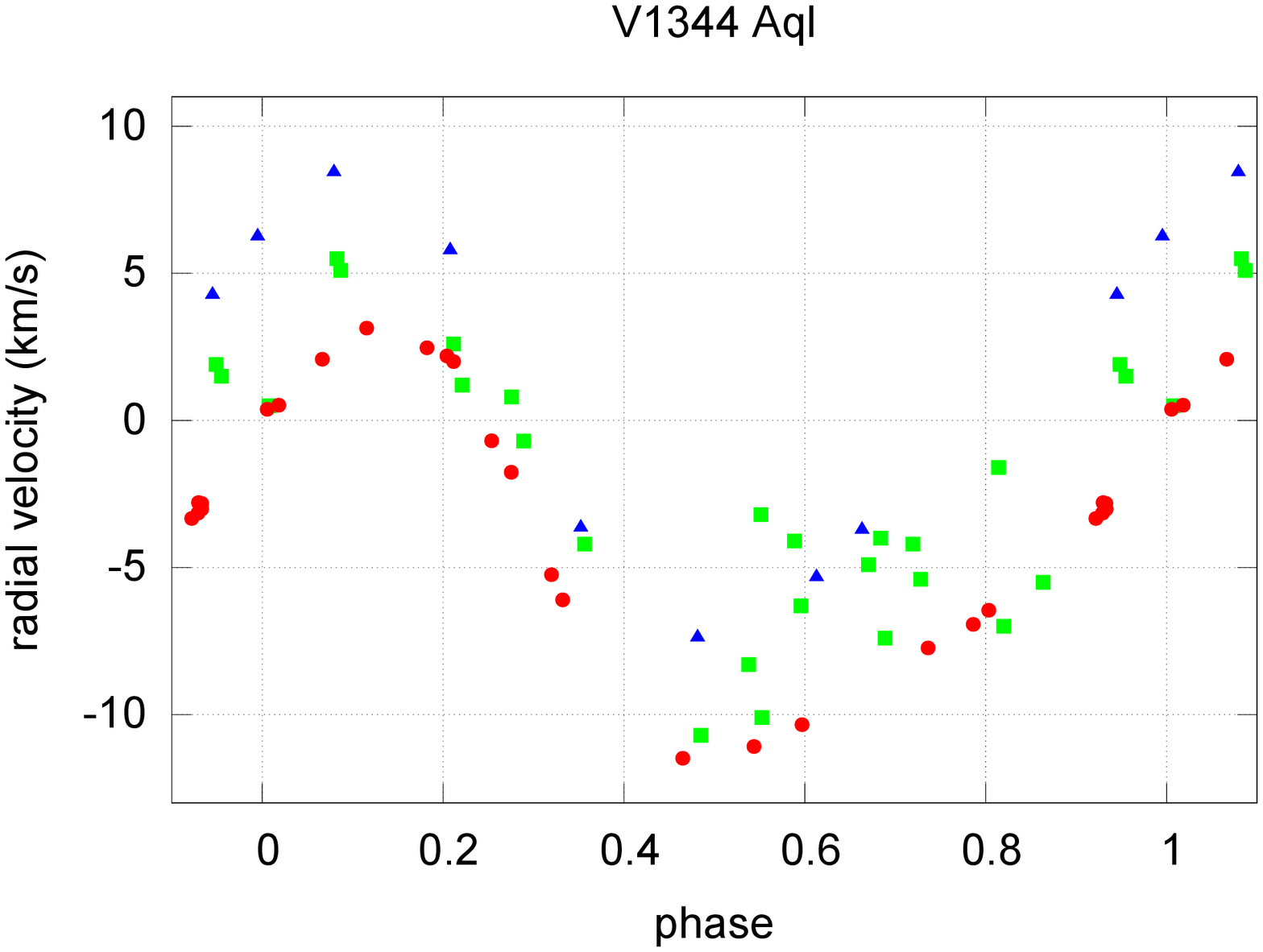}
}
\vspace*{-2.mm}
\caption{{\it Left:} An RV phase curve of GQ Ori based on very first medium-resolution echelle spectra taken in a Hungarian observatory ever. Data points are fitted with a Fourier polynomial of $n=5$. {\it Right:} An RV phase curve of V1344 Aql. Circles represent our new measurements, squares denote Balona's data (Balona, 1981), while Arellano Ferro's data (Arenalloi Ferro, 1984) are marked as triangles.}
\label{GQ-Ori-and-V1344-Aql}
\vspace*{-5.0ex}
\end{figure}

\acknowledgements
The authors would like to thank the Konkoly Observatory for using its 1m RCC telescope and its infrastructure. The project has been supported by the city of Szombathely under Agreement No.~S-11-1027. L.~Szabados thanks also ESTEC for supporting his work under Contract No.~4000106398/12/NL/KML.

\end{document}